# Different Stages of Wearable Health Tracking Adoption & Abandonment: A Survey Study and Analysis

Ahmed Fadhil
University of Trento
Trento, Italy
ahmed.fadhil@unitn.it

**ABSTRACT**
Health trackers are widely adopted to support users with daily health and wellness tracking. They can help increase steps taken, enhance sleeping pattern, improve healthy diet, and promote overall health. Despite the growth in the adoption of such technology, their reallife use is still questionable. While some users derive longterm value from their trackers, others face barriers to integrate it into their daily routine. Studies have analysed technical aspects of these barriers. In this study, we analyse the behavioural factors of discouragement and wearable abandonment strictly tied to user habits and living circumstances. A data analysis was conducted in two different studies, one with users posts about wearable sales and the other one was a survey analysis. The two studies were used to analyse the stages of wearable adoption, use and abandonment. Therefore, we mainly focused on users motives to get a wearable tracker and to post it for sale. We extracted insights about user motives, highlighted technology condition and limitations, and timeframe before abandonment. The findings revealed certain user behavioural pattern throughout the wearable use and abandonment.

**INTRODUCTION**
Commercial wearable health trackers have emerged as a way to track aspects of our daily activities. This may include tracking steps taken, sleeping patterns, daily diet tracking and heart rate monitoring. By observing lifelogs with these trackers, users are expected to improve their health status [20]. Health tracking bands are increasingly popular; however, they have high-levels of abandonment and little evidence exists to explain this obstacle. Although user optimism about the prospect of wearables, there exists a gap in the wearable functionalities and user expectations. Around 30% of users stop wearing their tracker within 6 months [11]. According to Gartner's 2016 consumer survey [15], user's boredom from using their wearable results in an abandonment rate of 2930%. The survey stated that smartwatch adoption is still in early adopter stage (10%), while fitness trackers have reached early mainstream (19%). According to the study, top reasons for abandonment include lack of perceived usefulness, boredom of use, or device break and malfunction. The study showed that people purchase smartwatch and fitness trackers for their own use, with 34% of fitness trackers and 26% of smartwatches received as gifts. Another study by Endeavour Partners in the US found that while one in 10 American adults own some form of activity tracker, half of them no longer use it. It is evident that wearable abandonment rate is higher relative to usage rate. Previous studies have analysed technical aspects and discussed that understanding nonuse of wearable technologies can provide insights that lead to future research design and implications [1, 27, 24, 6]. We took a step further analysis and looked at the user behavioural aspects of adoption/abandonment. This is accomplished in a twofold research analysis. The first part included reviewing posts on a second sale fitness trackers and smartwatches on Kijiji[1] for the Italian second sale wearable tracking technologies. A total of 484 posts about all wearables were reviewed from Kijiji. The review focused on analysing user motivation for post, health domain discussed within the post, adoption method and reasons for abandonment, usage frequency and duration before abandonment. We checked the wearable condition and technology limitations highlighted by users. All the analysed data were extracted from user posts and analysed to obtain accuracy with respect to user motivation and wearable features. We then conducted a survey with 26 participants all in Italy to test their adoption and abandonment reason, and the duration of use for their wearable. The two studies focused on measuring behavioural reasons on how the users adopt their tracker and reasons why they abandon them after some time. We additionally looked at the duration both groups used their wearable before abandonment and the main reason behind the abandonment. The wearables searched for in the first case were among the best wearables for 2017 according to PCMAG[2] review, a Tech, Gaming, Healthcare & Shopping review site. We considered Apple Watch, Fitbit Charge 2, Fitbit Surge, Garmin Forerunner, Mi Band 2, Misfit, Samsung Gear, and TomTom Spark as the selected wearables and smartwatches for the study. With this work we intend to make the following contributions: we provide a richer characterisation of the barriers that discourage or prevent, engagement with activity trackers over time; we highlight the main characteristics in terms of similarity and differences found in both studies, and finally we discuss the potential to enhance user engagement with such technologies.

---
[1] https://www.kijiji.it/
[2] https://www.pcmag.com/

## WEARABLE HEALTH TRACKING TECHNOLOGIES

Wearable health trackers are anticipated to proliferate in the market in the near future. Researchers have called tracking all aspects of one's daily life Lifelogging [28], quantified self [5], or personal informatics [22]. Wearables are capable of tracking steps, and other physiological information (e.g., heartbeat rate). Wearables use data stored to allow users gauge progress and gather incremental feedback. The data are visualised to enhance user's awareness about everyday activities and facilitate independent living and improve quality of life for citizens [25]. However, while some users derive value from their trackers for a long time, others find barriers to incorporating this technology into their routines. A study by Brandao et al., analysed the ex-users and current users of activity trackers and investigated factors of discouragement and reasons that could contribute to long-term adoption [4]. The findings suggested that long-term use is derived from the positive difference between the sense of usefulness and the effort necessary to maintain the continuous use of the devices. Much of the existing academic research on activity tracking has focused on use and behaviour change. Such studies have typically used devices supplied by researchers to provide appropriate designs for behaviour change. A review of 13 trackers found that they included 510 of 14 total behaviour change techniques identified from the research literature as potentially effective [12]. A work by Harrison et al.,
[12] investigated barriers to engagement with activity trackers. The study discusses implications for the design of activity tracking systems and how user customization could play a role. The sustained use of wearable technology is often the main challenge facing wearable technologies. Rather than describing the technological component employed in wearables, a work by Lee et al., [21] discusses the about the applications of current wearable technology from the sustainable perspective in the context of improving the quality of individual life, social impact, and social public interest. Successful and sustainable wearables will lead to positive changes for both individuals and societies overall.

Wearables are often touted as the greatest applications of the Internet of Things. This technology has the potential to transform our lifestyle by tracking health and exercise progress and bringing smartphones power to the wearer's wrist. Wearable devices can absorb extremely rich source of contextual information, such as conversation, location and gesture. A work by Billinghurst et al. [3] stated that wearables could sense handshakes, trigger face recognition and identify individuals. Although their benefit in terms of low-cost personalised healthcare, however, previous literature has identified several (non) technical and design related issues acting as barriers for wearable adoption in long-term [18, 6]. A work by Lam et al. [18] implemented a wireless wearable biosensors platform that uses biosensors and smartphone to measure heart rate, breathing rate, oxygen saturation, and estimate obstructive sleep apnea. The study addressed practical challenges in the design perspective of the platform. Another study by Jameson et al. [16] developed a mini wearable sensor device to enhance safety during ambulation for visually impaired users. The sensor warns them when they're about to hit obstacles at head level. The system emits an acoustic warning signals when a hazard is detected [16]. A similar work by Dakopoulos et al. [8] presented a survey among portable obstacle detection system as assistive technology for visually impaired people. The study analysed features and performance parameters of these wearables and provided a ranking as reference point of each wearable. Since wearables generate a vast amount of data, protecting this data is essential. A study by Pipada et al. [26] conducted a survey to gauge consumers concern about data security for wearable devices. The study explored possibility to accommodate these securities by the Technology Acceptance Model.

The majority of studies focus on technical aspects of wearable technologies and its correlation with user adoption [31, 2], little is known about individual characteristics and its correlation with wearable activity trackers adoption. Shih et al. [29] performed a six-week user study with 30 users using physical activity trackers embedded in clipon and smart watch physical devices. The study described user pattern implications, such as helping people be mindful of their physical activity tracker, to further articulate gender differences in use and adoption of wearable devices. To enhance user engagement in activity tracking, some wearables adapt exergames and gaming techniques. Wearable builders should engage users with incentives and gamification. Lindberg et al. [23] developed Running Othello 2 (RO2) exergame, in which players use a smartphone connected wrist band to compete in a board game enhanced with physical and pedagogical missions. The game uses inertial sensors and heart rate meter to detect physical activities of players. The findings revealed player's engagement with the game and identified challenges faced by users and how exergames with wearables could help. Moreover, to motivate users keep up with their personal activities, some wearables adapt motivational techniques, such as gamification and social recognition [7, 30]. The design space to leverage tracking data and persuade health related behaviour change involves issues and strategies to capture information, monitor progress, notify feedback, and provide social support [9, 17].

## METHODS AND STUDY DESIGN

Wearables still lack long-term user engagement due to factors, including wearable features, human perspective or behavioural aspects of the technology. A study by Lee et al.
[21] discussed the sustainable aspect of wearables in improving individual's quality of life, social impact, and social public interest. The notion is that by recoding information about user behaviours, such as physical activity or diet, the wearable can educate and motivate these towards better habits and health. The gap of recording information and changing behaviour is substantial, however little evidence suggests that they are bridging that gap. In what follows, we discuss the two studies we conducted to analyse the gap of wearable technology adoption/abandonment and user behaviour and motivation towards such technology.

**Post Data**
This part of the study included collecting posts from Kijiji about 8 of the best reviewed and highly rated wearables for 2017 on PCMAG technology review site. We picked wearables rated at least 4.5-5 stars, and strictly related to tracking aspects of health, such as diet, food journaling, sleeping patterns, physical activity and other health features, such as water and caffeine tracking.

Announcement Collection
We initially assumed that when a person posts wearable for sell, they're abandoning the technology and trying to sell it for no use/no benefits ...etc. Nonetheless, our investigations of the announcements and users motives behind their abandonment revealed other rationale behind the announcement, and not strictly related to abandonment. We reviewed the posts found on Kijiji between 10 and 31 January 2017. We searched for each selected wearable and smartwatch and we then conducted technical and behavioral analysis of the posts. The total number of announcements obtained was 158 for Apple Watch, 9 for Fitbit Charge 2, 15 for Fitbit Surge, 70 for Garmin Forerunner, 10 for Mi Band 2, 7 for Misfit, 208 for Samsung Gear, 7 for TomTom Spark, and 484 in total.

Posts Data Analysis
The data analysis focused on motivations for the announcement, the health motives discussed in the posts, the adoption method and reasons for the abandonment, the frequency and duration of use before abandonment, and the technical limitations highlighted by users together with technology condition. In Figure1 we show the pipeline followed to obtain and review the posts dataset with all the terms and features collected and deemed relevant to the posts.

**Survey Data**
This part of the study included preparing a questionnaire about wearable health tracking technologies and distributing it to participants. The questionnaire investigated reasons for wearable adoption and duration of use. It then asked users to highlight reasons for the abandonment and how they perceived the overall experience with the wearable device.

Survey Setup
Twenty-six participants (13 women and 13 men) took part in our survey, with an age range between 1832 (see Table1). Most of the participants have acquired a high school degree (n=21), however there were some with a graduate degree, with PhD Degree (n=3) and master's degree (n=2). Potential participants were recruited through the university network to take part. Our sample included both those who were currently tracking their activity, and those who had abandoned tracking. The sample also included those who have not used any wearable tracking devices. Participants tracked with a range of wearables, including Garmin Vivofit, Mi Fit and LG G Watch. The most experienced participant reported having used a tracker for more than a year, while the least experienced had used it for only 1 day. The rest reported using their tracker for 1 month, 2 months, 6 months and 1 year. Participants also reported using other personal informatics tools in addition to activity trackers, smartwatches, such as Samsung Smartwatch, Motorola Moto 360 Smartwatch and Google Glass.

| Participants Demographics | | |
|---|---|---|
| Gender | Frequency | Age Range |
| Male | 13 | 1932 |
| Female | 13 | 1827 |

Table 1. Participants Demographics.

Survey Data Analysis
The survey analysis focused on investigating the method of adopting the tracker, whether it's through purchase, gift or win. It also analysed the usage frequency and duration and reasons for the abandonment. In Figure2 we show the survey structure and flow.

**POSTS AND SURVEY DATA MAPPING**
The two datasets reference two different stages at which a wearable is abandoned. For example, while the survey conducted to understand whether users have any wearable, and motives behind their decision to use a tracker or abandon it after a certain time, the post data mostly focuses on the last stage of wearable use, that is abandoning the tracker. However, one should notice that in some cases even though a user posts to sell their wearable, their motivation is rather different than abandonment. To illustrate, if the user is selling for an upgrade or double gift reason, this is different than selling for no use and abandonment.

**Adoption Motives**
This refers to the way in which the users obtained the wearable tracker. Most Kijiji posts mentioned the direct purchase and gifts reviewed as the main adoption method, whereas survey respondents provided that they reviewed it as a gift, bought for curiosity, and to promote their health. A big portion of the posts on Kijiji data didn't mention clearly the adoption mechanism. In Table2, we highlight the adoption mapping for both the posts and survey data.

| Adoption Motives | | |
|---|---|---|
|  | Kijiji posts | User survey |
| Curiosity |  | x |
| Useful functions |  | x |
| Comfort |  | x |
| Tracking health |  | x |
| Promoting Health |  | x |
| Gift | x | x |
| Direct purchase | x | x |

Table 2. Posts-Survey Adoption Motives Mapping.

**Usage Duration**
With the usage duration we measure the period a user adopts a wearable and the period they continue using it or stop using it at some point. Usage duration given by the posts and survey referred to the period each technology was used. This is a useful predictor to estimate the timeframe in which a user decides to abandon a given technology or continue using it. The usage duration from both studies ranged from days, weeks, months, to years. The timeframe range was decided after reviewing the post and survey reviews, where the user mentioned the duration, they had the technology before they posted it for sell or they stopped using it. Some examples include, "I used it for 1 day just to test it" for a daily duration, "I used it for only one week" for a weekly duration, "I used it for 2 months only" for monthly duration and "It works perfectly it has a year of life" for yearly durations. In Table3, we highlight the usage duration of the trackers before abandonment as it was extracted from both datasets and perform a mapping on the common data points.

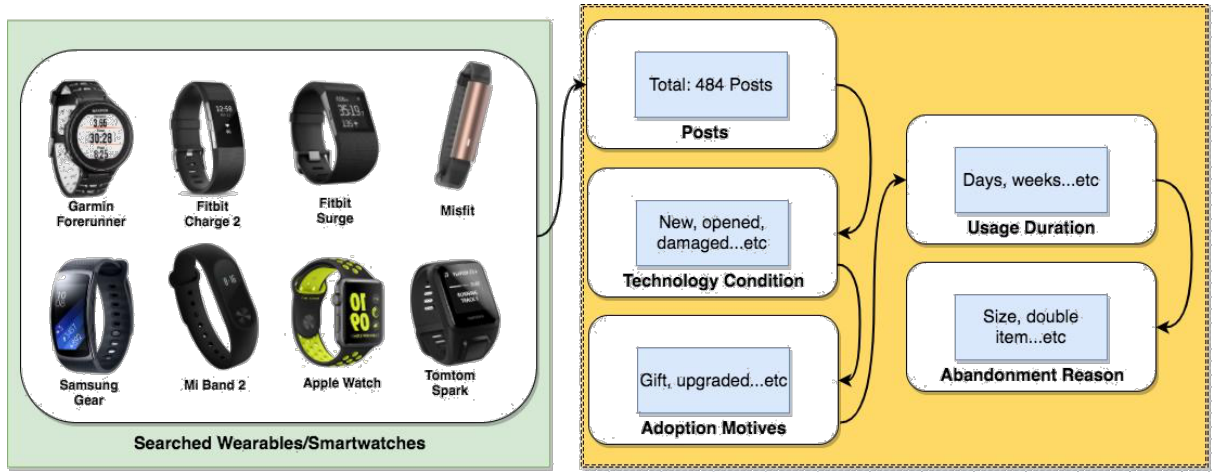

Figure 1. Posts Collection and Analysis Pipeline.

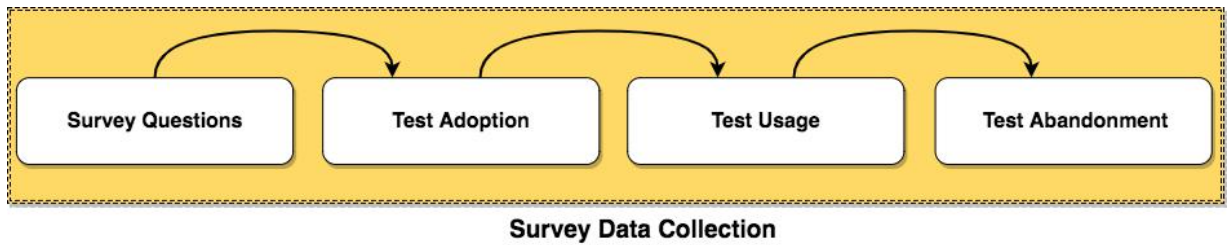

Figure 2. Survey Structure and Flow.

| Usage Duration | | |
|---|---|---|
| | Kijiji posts | User survey |
| 1 Day | | x |
| Days | x | x |
| Weeks | x | |
| 1 Month | x | x |
| Months | x | x |
| 1 Year | x | x |
| >1 Year | x | x |

Table 3. Posts-Survey Usage Duration Mapping.

**Abandonment Reason**

To investigate the use of the wearable, we had to also analyse the abandonment reasons highlighted by both the posts and the survey data. For that, we checked each post for reasons behind the abandonment and included this as a question in the survey to understand the various motives users decide to abandon their wearable. Most of the abandonments on Kijiji posts were due to change in accessories, upgrading model, gift, user preferences, perceived usefulness, wrong purchase, tech curiosity, double items, compatibility issues, feature requirements and size issues. Similar patterns were found when we analysed the survey data. For example, users were selling their wearable because they received it as a gift, perceived usefulness, wrong purchase, tech curiosity. In addition, other users highlighted that perceived complexity of use, and inefficiency with the battery life was the reason they stop wearing or totally abandoned their tracker. Although abandoning the technology is perceived negatively, however we found cases where the user was selling their device for upgrade or double gift, in which case it's not considered as abandonment. In such cases the seller is still in the context of use, although he is switching or selling extra devices. In Table4, we highlight reasons behind technology abandonment as was given by both analyses.

**FINDINGS**

The findings revealed common abandonment motives, although the adoption motive was for different purposes, including physical activity, sleep pattern and food tracking. In both of our analysis, users abandoned their wearables for motives other than perceived lack of utility or borne out of frustration and boredom with the device. For example, the analysis revealed abandonments for reasons related to double gifts received by users, upgrading to newer version or different models, or selling for wrong purchase. The study also revealed that majority of wearables featured at least three health and wellness elements, including physical activity, sleep and diet tracking. The frequency of use before abandonment varied from 1 day to more than a year.

| Abandonment Motives | | |
|---|---|---|
| | Kijiji posts | User survey |
| Accessories Change | x | |
| Upgrading Model | x | |
| Gift | x | x |
| User Preferences | x | |
| Perceived Usefulness | x | x |
| Wrong Purchase | x | x |
| Tech Curiosity | x | x |
| Double Items | x | |
| Compatibility Issues | x | |
| Feature Requirements | x | |
| Size Issues | x | |
| Perceived Complexity | | x |
| Battery life Efficiency | | x |
| Priority Change | | x |

Table 4. Posts-Survey Abandonment Reason Mapping.

Based on the review there exists patterns in user behaviour towards the announcement and reasons beyond the abandonment of the technology. For example, a motivation to sell a wearable was because of double gifts, which often times resulted in having two items of the same brand. This is not strictly a reason for abandonment, but rather most probably a continues use. Since the user didn't highlight any intention to abandon the technology, but rather they had other motives behind their decision. Other trends in the data were related to the health domain covered by the technology and health reasons mentioned by users for the abandonment. For example, some decided to post their wearable because they had a health status change, as in "I sell because I can't use it due to health related issue", or because of a change in their lifestyle, as in "I lost weight and achieved my goal, I don't use it anymore" and "I bought a bike and I can't use it to track my steps". The usage duration was spread and ranged from 1 day, as in "Bought at the day one, its new" to few years, as in "2 years of life, but taken good care of". There were many similarity patterns among the posts obtained from both Kijiji and the survey data. Interestingly, some (non) technical were obtained from the posts provided insights on why they abandon the wearable and what are the associated limitations led to the posts that were highlighted by the users. For example, some reasons were related to the bracelet size for the wearable, as in "The band is too small for me", or for a feature requirements, as in "The band is not water proof", or simply for user's personal preferences, such as color, as in "I sell because I bought a black one". We consider these findings in our design recommendations for wearable technologies that focuses on sustained evolving adoption. Our work contributes to existing research on wearable technologies and user's engagement by analysing patterns in user engagement or frustration with the wearables. These will help advance the research on wearable technologies and evolving adoption rates.

**Stages of Wearable Usage**

The study analysed two different wearable usage data through the survey and the posts. The main goal was to define a pipeline describing the different stages a user may decides to get a wearable tracker, user it, and either continue the usage or abandon it after some time.

Idle

At this stage the user has no wearable tracker and has no intention to get one anytime soon. This step describes those users who have no interest in getting/using any wearable health tracker. This was evident from the survey data where users asked why they don't use a wearable (see Table5 for user responses). According to user reply, the majority don't have a wearable because they are not interested or its too expensive.

| Why wouldn't you own a wearable? | |
|---|---|
| Reasons | Frequency |
| Too Expensive | 6 |
| Not Interested | 15 |
| Unknown | 5 |

Table 5. Reasons For Not Owning a Wearable.

Pre-Adoption

This stage refers to the period where the user thinks to get a wearable tracker, but he/she isn't ready to act immediately. This could be correlated to user's overall awareness of their lifestyle. For example, if the user follows a poor lifestyle and decides to change his current lifestyle, then he/she might think to get a wearable tracker to improve aspects of their lifestyle (e.g., lose weight, burn calories, improve sleep, etc). In such cases the user might decide to either shift to the adoption state or return to the idle state.

Adoption

At this stage the user decides to take action (i.e., get a wearable tracker). The action depends on what the user plans to use the wearable for. The adoption method may also play a crucial role in how the user perceives the wearable tracker. For example, if the user decides to get a wearable tracker to lose weight is very different than when they receive the wearable as a gift. This stage was analysed by both of our studies, through a question in the survey and analysing users posts in the wearable post analysis. As we mentioned in Table2 above, the adoption motives may vary depending on user's lifestyle, health awareness of the user or their family member. The adoption motives comparison between the two data revealed that users provided more responses to the reason for wearable adoption, whereas the post review data revealed only two reasons, namely gifts and direct purchase as the main reasons for wearable adoption (see Table6 and Table7 for the adoption motives by the survey and the posts, respectively).

Post-Adoption

Users at this stage are familiar with wearable trackers and use it on daily basis. This stage is critical to decide if the user will continue using the device or abandon it at a certain time. Both studies showed a range of usage duration from 1 day to more than a year. Why some users choose to wear the tracker for one day and why others continued wearing it is an open research question and interesting research investigation to analyse user motives towards wearable tracker use. Perhaps this stage is the most important one in determining wearable tracker adoption and abandonment. In Table8 we list the usage duration after adopting a wearable health tracker, and in Table9 we list the wearables considered for the post analysis study with their usage duration before posting the sale.

| The motives for using a wearable? | |
|---|---|
| Responds | Frequency |
| Curiosity | 1 |
| Useful functions | 1 |
| Comfort | 2 |
| Have an alternative watch and see my steps | 1 |
| Healthy life | 2 |
| My dad bought it to me | 1 |
| Testing a new wearable technology | 1 |

Table 6. Adoption Motives By Survey.

| Adoption Motives | |
|---|---|
| Kijiji Posts | |
| Gift | 34 posts |
| Purchased | 213 posts |
| Unspecified | 237 posts |

Table 7. Adoption Motives By Posts.

| How long have you had your wearable device? | |
|---|---|
| Period | Frequency |
| Never | 14 |
| 1 Day | 1 |
| 1 Month | 1 |
| 2 Months | 1 |
| 6 Months | 2 |
| 1 Year | 1 |
| More than a year | 6 |

Table 8. Usage Frequency Before Abandonment Survey.

Abandonment
At this stage the user stops using/wearing the tracker after a usage duration that may vary from few days to few years. This is often followed by first leaving the tracker at home and then decide to post it on a secondhand store. We extracted meaningful insights from both studies to further analyse motives behind user decision to stop using the tracker. One interesting fact we found from the post data analysis was that people may post a wearable for sale, but not necessarily to abandon it. For example, in cases where the user decides to upgrade to a newer model or a newer tracker, or sells because of double gifts, the user is still engaged with the wearable and he/she is still in the postadoption state. We do not consider such cases as abandonment. We list below in Table10 the reasons given by survey users about their abandonment, and in Table11 the information extracted from the post data about the reasons for their posts.

We constructed a list of keywords from the analysis that better describes what users have highlighted. These keywords were color change, upgrading model, gift, user preferences, perceived usefulness, wrong purchase, tech curiosity, double items, compatibility issues, feature requirements, and size issues. Based on these keywords, we obtained the frequency of each limitations as given by the posts.

However, this process of adopting a wearable, using it and then abandoning it after some time involves also individuals recycle between stages. This means that a user may return from abandonment into postadoption stage and the later may either proceed or fall into the abandonment stage again. To confirm this point, we included an extra question in our survey to analyse whether a user who abandoned their tracker will be willing to go back and reuse the tracker again (see Table12).

Although there is a chance the user might return back to use their wearable, however, according to our study, the possibility a user might go back to reuse their wearable is minimal. In Figure3 we show the stages an individual may go through to get and start using a wearable and either continue using it or abandon it after sometimes.

**Wearable Adoption Abandonment Comparison**
Based on the two data analysis and the stages of wearable adoption and abandonment, we compare the two datasets and mapped between the reasons to use activity tracker and reasons to abandon or sell it. This is important to understand the two stages of wearable use to decrease the relapse. In Table13 we highlight the reasons users motioned to own or sell a tracker.

**LIMITATIONS**
We performed two separate analyses to investigate the stages of wearable technology adoption and abandonment. We performed a survey study to investigate why users decide not/to have a wearable and for how long. The second part focused on reviewing the announcements and extracted wearables and smartwatches related posts. In the case of Kijiji posts, we found that the majority of users perceived the technology as useless (38 posts). Interestingly, this was followed by upgrading the model (19 posts). We believe that uselessness and upgrading aspects of limitations are very different, since it is very different from the uselessness aspect; the first one is resulting in abandoning the technology, whereas the second one is continuing the use. In the case of survey responses, we also found that the majority of responses abandoned the device due to no intention to use (15 responses). The announcements mostly mentioned perceived uselessness of the technology. For example, a user mentioned "I sell it because I cannot use it". However, other cases included selling the old model to upgrade to newer models. For example, "I sell by switching to 2 'series because of the need for GPS". Other less frequent cases included color change (e.g., "I sell because I bought a black one"); gift, user preferences (e.g., "I sell because it's a bad gift"); wrong purchase (e.g., "I bought for incorrect steel bracelet purchase"); tech curiosity (e.g., "Purchased for curiosity"); double items (e.g., "because I already have one like that and that's enough"); compatibility issues (e.g., "I sell because it is not compatible with my notes"); feature requirements (e.g., "The band is not water proof"); and size issues (e.g., "The band is too small for me"). In the case of survey responds, the users mostly mentioned perceived uselessness of the technology. For example, a responder mentioned "Because I don't find it useful". Other cases were related to the efficiency of the wearable and user experience, as in "Because charging it once a day was so boring", or "It broke". Several limitations are associated with this study, partially related to data collection and the analysis. There are few items reported as damaged, this is since we believe seller wants to promote their item to be sold. Moreover, the majority of posts did not highlight motivation, condition, technical limitation behind the abandonment. This can introduce a bias in the data analysis, where the seller mostly describes positive aspects of the technology and minimizing negative details, since there motivated to maximise profits and quickly sell their item. Moreover, posts that didn't describe negative aspects of their experience with the wearable might question their real motive behind the abandonment. In addition, having some unclear and ambiguous posts made it hard to determine the actual motive behind the posts. In the case of the survey, one limitation is the small user group compared to the posts. However, still collecting this data helped us to answer interesting questions about the preadoption, adoption, and postadoption stages. Some abandonment cases were perceived as positive abandonment of the device, similarly to the study by Grudin et al. [10], we found some cases where the abandonment was due to seller's desire to upgrade to more advanced trackers, or it was a gift, or even related to size issues. These causes aren't necessarily abandonment of the tracker, since there is no indication of a drop in user intention to reuse the application. The study has shown other usage patterns and motives for abandoning the wearable. For example, usage duration before abandonment have shown the timeframe users spend before selling the device. Although this is not sufficient to conclude their actual motives for abandonment, however we believe studying the relation between usage time and user decision and link them with the abandonment will help understand usage pattern among different wearable users. This is outside the scope of this study; however, we invite researchers to conduct some studies on this issue. Technology limitations highlighted by users was another interesting pattern about user motives for abandonment. We have obtained both technical (e.g., I sell by switching to 2 series because of the need for GPS) and nontechnical (e.g., Purchased for curiosity) indications about the abandonment.

| Usage Frequency Before Abandonment | | | | | | | | |
|---|---|---|---|---|---|---|---|---|
| Smartwatches/Wearables | Never Used | Immediate | Day(s) | Week(s) | Month(s) | Year(s) | Unclear | Not reported |
| Apple Watch | 27 | 14 | 3 | 2 | 6 | 3 | 53 | 72 |
| Fitbit Charge 2 | 2 | 2 | 1 | | 1 | | 1 | 2 |
| Fitbit Surge | 5 | 4 | | | | | 2 | 4 |
| Garmin Forerunner | 9 | 8 | | | 3 | 3 | 6 | 32 |
| Mi band 2 | 3 | 1 | | | 2 | | | 4 |
| Misfit | 3 | 1 | | | | | | 3 |
| Samsung Gear | 110 | 21 | | | 4 | | 11 | 49 |
| TomTom Spark | 1 | 1 | | | | | 2 | 3 |

Table 9. Usage Frequency Before Abandonment Posts.

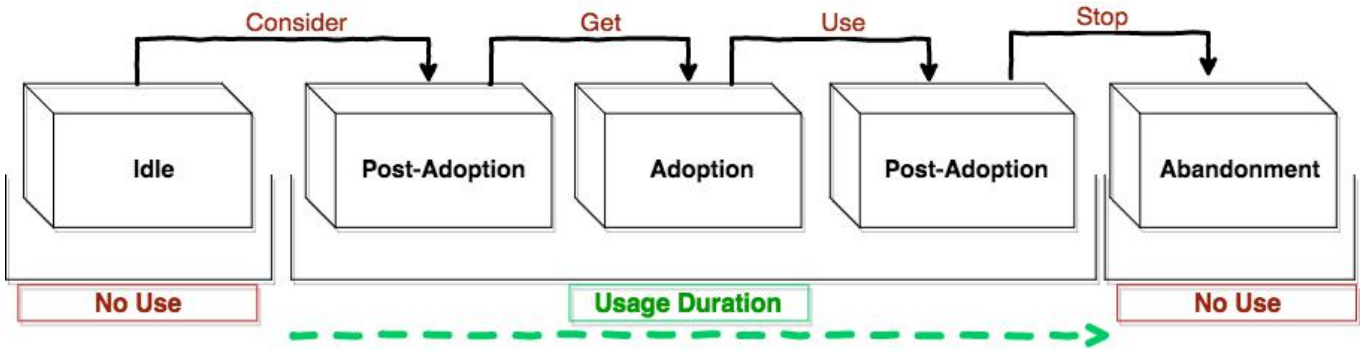

Figure 3. Stages of Wearable Adoption and Abandonment.

| Why you stopped wearing your device? | |
|---|---|
| Reasons | Frequency |
| It broke | 1 |
| I don't use it as much as I used to | 1 |
| Not interested | 1 |
| I don't need it anymore | 1 |
| Because I don't find it useful | 1 |
| Because it's too complicated to use | 1 |
| Because I don't like wearable tech devices even if I have one | 1 |
| Because charging it once a day was so boring | 1 |
| Its tedious | 1 |

Table 10. Abandonment of the Wearable Tracker Survey.

| Abandonment Reason | | |
|---|---|---|
| Kijiji Posts | | |
| Accessory change | Immediate | Upgrading model |
| User preferences | Perceived usefulness | Wrong purchase |
| Double items | Compatibility issues | Feature requirements |
| Tech curiosity | Size issues | Gift |

Table 11. Abandonment of the Wearable Tracker Posts.

| Would you go back and wear it again? | |
|---|---|
| Responds | Frequency |
| Yes | 5 |
| No | 17 |
| No Responds | 4 |

Table 12. The Recycling between Stages.

| Usage/Abandonment Comparison ||
| Reasons to adopt | Reasons to abandon |
| --- | --- |
| My dad bought it to me | Selling because I got it as gift |
| Using it to measure my steps | I don't use it anymore |
| Bought it for curiosity | I sell if after a week of use |
| Got it to measure my distance | Selling because I can't use it for health reasons |
| Purchased it to measure my steps taken | Selling it because I am using a bike and have no use for it |
| Bought it for its comfort | Selling because I don't want to charge it everyday |

Table 13. Wearable Adoption Abandonment Comparison.

**DISCUSSION AND FUTURE INSIGHTS**
There is a grim picture regarding wearable technologies and their ability to enhance long-term engagement to achieve meaningful goals or enact changes in user's health behaviours [19, 14, 13]. Recent studies have reported a high rate of wearable abandonments by users [10, 6]. These studies questioned the core functionality of wearable technologies and concluded that either the overall vision for these technologies is misplaced, their design is deeply flawed, or both. However, our study shows that there exist many reasons for the abandonment beyond the technical and design aspects. User intention and wearable capabilities, flexibility of the wearable to adapt and user habits are all factors affecting user engagement with the wearable in the long-term. Using persuasive elements in the wearable design should clearly define the role of the persuasive technology in sustaining user adherence. For example, whether the persuasive tool will be permanent and always present in the system or it will be temporary and will be off after the user achieves a certain level. In addition, the role of behaviour change theories in wearable technologies often takes a static view of the user and do not account for changes in users' circumstances. Adding more flexibility to the wearables could adapt to user's changing circumstances. There is a need to account for new streams of information, flexibility with user grow, and focus on design aesthetics when making wearable technologies for health. In summary, current research focuses on technical and device related limitation and tries to account for user's low adoption rate by further investigating these points. Fewer studies have focused on user habit and behaviour oriented abandonment cases in wearable and smartwatch technology.

Investigating reasons behind user abandonment of wearables through real post investigation or survey questionnaire is an important step towards finding insights about technical limitations led to the abandonment. However, focusing only on technical limitations is still not enough to fully understand user motives for the abandonment. Besides analysing technical limitations that led to the announcements, we particularly focused on user behaviour and intention for the abandonment through a thorough analysis of all the posts and survey data. The findings revealed some correlation pattern between wearable analysis from technical perspective and users' tone from behavioural perspective. Our goal was to provide insights on how to build enticing wearables, and to enhance their long-term use. The finding provided us with insights to answer our question: How to increase wearable devices adoption rate?

**CONCLUSION**
There is a big disruption with personal health tracking technologies, as they are rapidly adopted into mainstream culture and have sparked an explosion of interest in tracking various aspects of health. However, these technologies suffer from being largely abandoned in the long-term. Current research investigating this issue focuses on technical aspects of the abandonment related to wearable features and functionalities. While this is necessary to improve the quality of services offered. However, behavioural aspect plays a great role in understanding motives behind this abandonment, which was relatively unexplored. In this paper we present the findings from two studies focused on characterising barriers that affect engagement with activity trackers. The first study analysed posts about wearable sale on Kijiji. Where we investigated technical limitations of the abandonment, however, our focus was on the behavioural aspect. For that, an iterative analysis was conducted on all posts and we extracted useful insights and patterns of abandonment from all the posts.

The second part focused on a survey questionnaire to study user's motives behind wearable adoption and abandonment. The empirical analysis and findings from both studies revealed cases of abandonment, technology limitations highlighted by users, and user intention to abandon the device. In many cases users abandoned their device for personal reasons, and not necessarily due to technical limitations. Understanding user's circumstances, their intend of use, and what the wearable offers could enhance the design and long-term adoption of such technology. There is a need for foundational guidance for future wearable design and development and UX guidelines that incorporate selfefficacy and serves consumer healthcare engagement. The insights from our findings suggest more research on design aspect, theoretical foundation for user behaviour, motives and expectations from wearables.